\documentclass[prb,showpacs,floatfix,amsmath,amssymb,superscriptaddress]{revtex4}
\usepackage[dvips]{graphicx}
\usepackage{latexsym}
\usepackage{graphicx}
\usepackage{times}
\usepackage{amsmath}
\usepackage{dcolumn}
\usepackage{subfigure}
\usepackage{latexsym,amsmath,amssymb,bm,euscript}
\usepackage{amsfonts}
\usepackage{hyperref}

\newcommand{\ket}[1]{| #1 \rangle}
\newcommand{\bra}[1]{\langle #1 |}
\newcommand{\dirac}[2]{\langle #1 | #2 \rangle}

\newcommand{\be}{\begin{equation}}
\newcommand{\ee}{\end{equation}}

\begin{document}

\title{Bulk metals with helical surface states} 
\author{Doron L. Bergman, Gil Refael}
\affiliation{Physics Department, California Institute of Technology,
  MC 114-36, 1200 E. California Blvd., Pasadena, CA 91125}

\date{\today} 
 
\begin{abstract}
In the flurry of experiments looking for topological insulator materials, it has been recently discovered that
some bulk metals very close to topological insulator electronic states, support the same topological surface states
that are the defining characteristic of the topological insulator. First observed in
spin-polarized ARPES in Sb (D. Hsieh et al. Science 323, 919 (2009)),  
the helical surface states in the metallic systems appear to be robust to at least mild 
disorder. We present here a theoretical investigation of the nature of these ``helical metals'' - bulk metals 
with helical surface states. We explore how the surface and bulk
states can mix, in both clean and disordered systems. Using the Fano
model, we discover that in a clean system, the helical surface states are \emph{not} simply absorbed by
hybridization with a non-topological parasitic metallic band.
Instead, they are pushed away from overlapping in momentum and energy with the bulk states,
leaving behind a finite-lifetime surface resonance in the bulk energy band.
Furthermore, the hybridization may lead in some cases to multiplied surface state bands, in all cases retaining the helical characteristic.
Weak disorder leads to very similar effects - surface states are pushed away from the energy bandwidth of the
bulk, leaving behind a finite-lifetime surface resonance in place of the original surface states.
\end{abstract} 
 
\maketitle 

\section{Introduction}
\label{intro}

The prediction and subsequent discovery of bulk topological
insulators (TI) has galvanized the condensed matter community, not in the
least because of the unique physics of the
protected edge states on their surface\cite{Kane:2005A,Bernevig:2006,Kane:2005B,Kane:Science2006,Bernevig:prl2006,Wu:2006,
Konig:Science2007,Fu:2007,Moore:2007,Teo:2008,Kane:2008,Schnyder:2008,Qi:2008,Hsieh:2008,Fu:2008,
Zhang:nature2009,Buttiker:Science2009,Xia:2009,Zhang:prb2009,Roushan:2009,
Ran:2009,Roy:2009A,Roy:2009B,Schnyder:2009,Chen:2009,Essin:2009,Moore:2009}. 
Many materials adjacent to TI's, however, are bulk
metals with the same helical surface states, e.g.,
Sb\cite{Hsieh:2009,Hsieh:2010}, Bi$_{0.91}$Sb$_{0.09}$\cite{Taskin:2009,Taskin:2010}, 
Bi$_{2-x}$Mn$_x$Te$_3$\cite{Hor:2010}, and even undoped
Bi$_{2}$Se$_{3}$\cite{Analytis:2010,Eto:2010}, which initially was thought to be a wide band-gap TI, but now seems to have a small 
but measurable bulk Fermi surface. In such materials one would naively expect
that a bulk Fermi surface would simply swallow the surface states. 
Nevertheless, the same helical surface states of the TI phase, continue to
appear in them, now coexisting with a bulk Fermi surface. We suspect
that more examples will emerge, especially since the helical surface states are less sensitive to disorder than ordinary (non-helical) 
surface states (of the likes of those found on the
$\langle 111 \rangle$ surfaces of Cu), as has been already suggested by surface-doping Sb with Pottasium\cite{Hsieh:2009}.

\begin{figure}
	\centering			
			\includegraphics[width=3.0in]{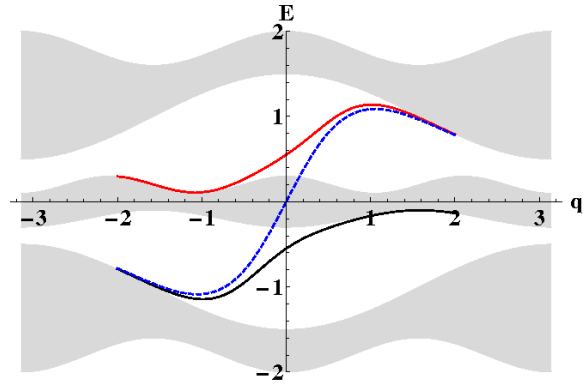}
	\caption{An energy-momentum sketch of the fate of surface states when they
	overlap with a parasitic metallic bulk band. The solid (light green) regions
	mark the bulk density of states, the top and bottom bands being
	bulk states of a topological insulator, and a parasitic metallic band in the
	middle. The original surface state branch (dashed line) overlaps with the middle (metallic) band.
	The surface state branch hybridizes with the bulk, leaving a diffuse resonance in the bulk band 
	energy range, as well as two sharp edge-state branches appearing above and below the parasitic metal energies.
	This demonstrates that the topological surface states, rather than being obliterated, are simply pushed away from overlapping with the bulk.}
	\label{fig:exiled-states}
\end{figure}

The prevalence of metallic systems with surviving helical surface states led us to
ask: what exactly happens when a surface-state, which is the result of a TI bulk,
is allowed to hybridize with a 'parasitic' non-topological metallic
band. As we shall see below, not only do the helical edge states
survive, but they can actually multiply. 
When mixed, the surface states are simply pushed away from overlapping in energy and momentum with the bulk states.
If the surface and bulk do not overlap to begin with, the mixing is not effective, and the surface states are only slightly modified.
In those areas of energy and momentum overlap between the bulk and surface states, \emph{new ``exiled'' surface states} appear above and below the
confines of the metallic band, and in place of the original surface state a ``ghost'' surface resonance remains, with a finite (and often 
very short) lifetime. Thus in one momentum value, we may see ARPES
signatures of two surface states at energies above, and below,
the metallic energy range (see Fig.~\ref{fig:exiled-states}), in addition to a surface resonance within the bulk metallic band.   
The exiled states, as well as the remaining finite-life-time {\it ghost} 
resonances will retain the odd-number distinction of surface bands, characteristic of
helical surface states, comprising an odd number of Kramer's pairs of bands in 2D, and an odd number of Dirac cones in 3D.

%
%
%

In this manuscript, we explore the
metal vs. edge-state struggle by first constructing 2d examples
where a metallic band appears at the same energy as an edge state, and
numerically investigating its emerging spectral structure. Next, we
approach the problem analytically by constructing a generic model for
a helical surface state interacting with a bulk metal, based on the
Fano model\cite{Fano:1961}. From both approaches, the generic picture of
a {\it helical metal} arises, as is summarized in Fig. \ref{fig:exiled-states}.

\section{Construction of 2d helical metals}
\label{numerics}


We begin our study by attempting to construct disorder free
theoretical models in 2d where helical edge states coexist with a bulk
metal in both momentum and energy. Such helical metals can be achieved in at least two different ways.
One possible construction is to take a model of a TI, add a new (initially decoupled) partially filled band,
an then mix it with the TI bands.
A second possible construction is to add a momentum-dependent chemical
potential without adding additional degrees of freedom, such that 
the gap closes somewhere in the Brillouin zone (BZ). This can always be achieved in a tight binding model by adding appropriate hopping 
terms that produce purely diagonal terms in the multiband hopping model. 
We will consider an explicit example of the former, demonstrating the
construction in 2d. An explicit example of the latter construction is relegated to the supplementary material (see Section~\ref{2band_model}),
as the main physical features of it are no different than in the one
example we show here). Our 2d examples can easily be generalized for models of topological insulators in any dimension.

\begin{figure}
	\centering
		\includegraphics[width=1.5in]{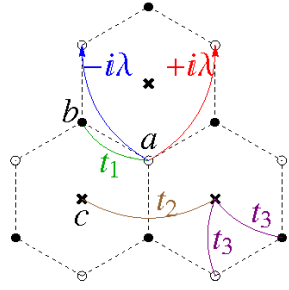}
	\caption{A two dimensional toy model for the investigation of a helical metal. The model is based on the Kane-Mele 
	model\cite{Kane:2005A,Kane:2005B} on the honeycomb lattice, which is denoted by dashed lines. The Kane-Mele model
	consists of nearest-neighbor hopping $t_1$ (green) between the two ($a,b$) sublattices of the honeycomb lattice 
	(denoted by empty and filled circles respectively), and a complex second neighbor hopping (a spin orbit coupling term)
	with opposite sign when clockwise ($-i \lambda$, blue) and counterclockwise ($+i \lambda$, red),
	as indicated by the curved arrows in the figure.
	In addition to sites of the honeycomb lattice, we include a new set of sites ($c$) at the centers 
	of the hexagonal plaquettes of the honeycomb lattice, denoted by cross marks. The $c$-sites form a triangular sublattice, 
	and nearest neighbor hopping between them $t_2$ (brown) forms a metallic band. 
	To explore the interplay between the topological insulators helical surface states and the bulk metallic band we mix the 
	two systems by allowing hopping between the $c$-sites and the honeycomb lattice sites, $t_3$ (purple). 
	}
	\label{fig:model_conventions}
\end{figure}

\begin{figure}
	\centering			
		\subfigure[ Zigzag edge, $t_3 = 0$]{
			\includegraphics[width=3.0in]{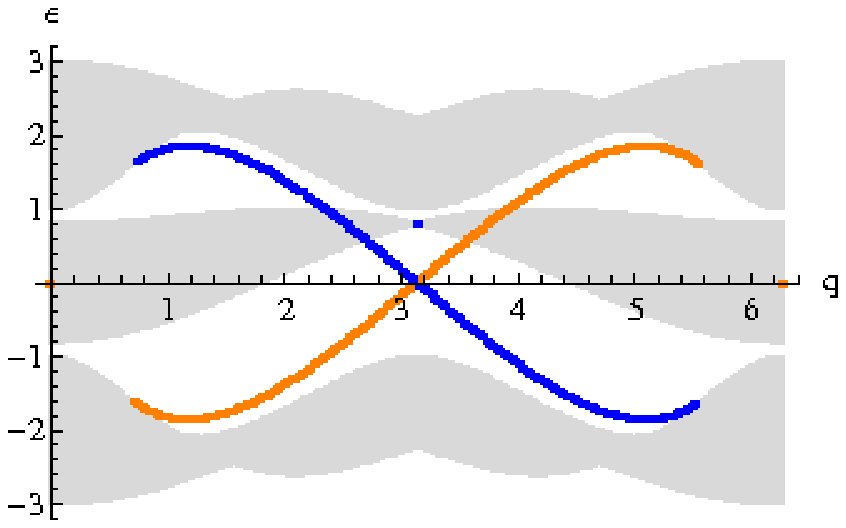}
			\label{fig:subfig1}
		}
		\subfigure[ Armchair edge, $t_3 = 0$]{
			\includegraphics[width=3.0in]{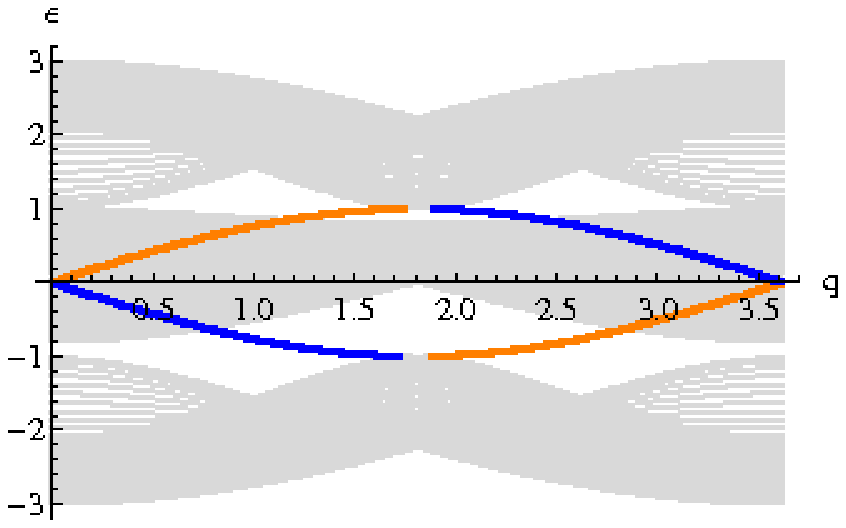}
			\label{fig:subfig2}
		}		
		\subfigure[ Zigzag edge, $t_3 = 0.03$]{
			\includegraphics[width=3.0in]{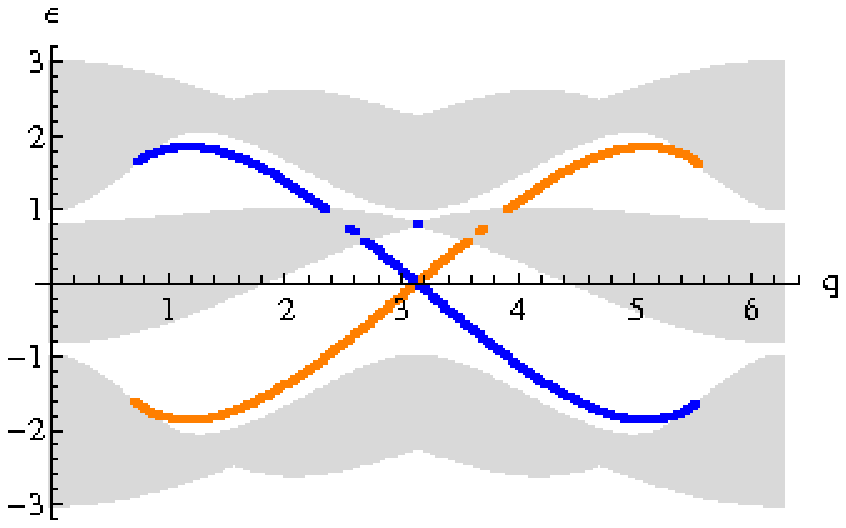}
			\label{fig:subfig3}
		}
		\subfigure[ Armchair edge, $t_3 = 0.03$]{
			\includegraphics[width=3.0in]{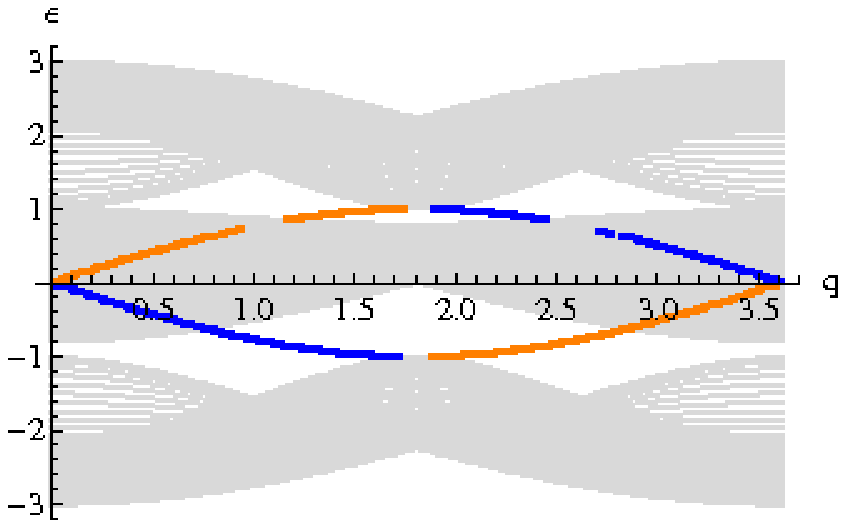}
			\label{fig:subfig4}
		}
		\subfigure[ Zigzag edge, $t_3 = 0.3$]{
			\includegraphics[width=3.0in]{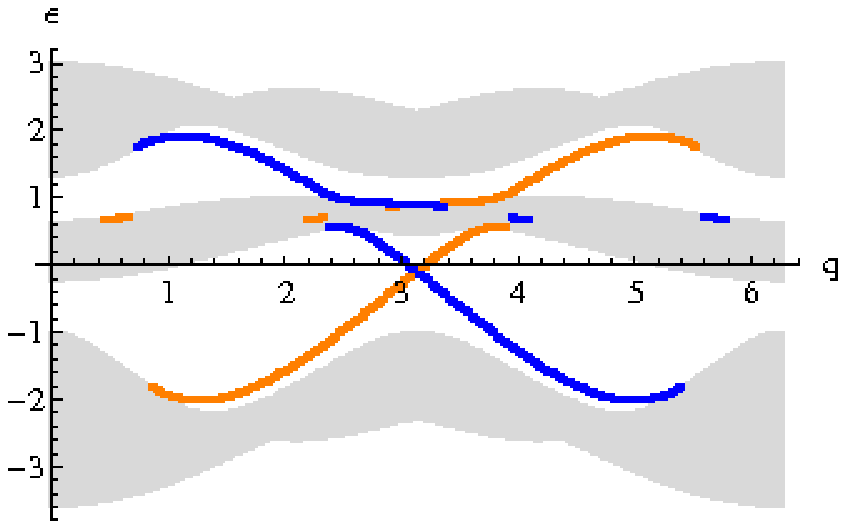}
			\label{fig:subfig5}
		}			
		\subfigure[ Armchair edge, $t_3 = 0.3$]{
			\includegraphics[width=3.0in]{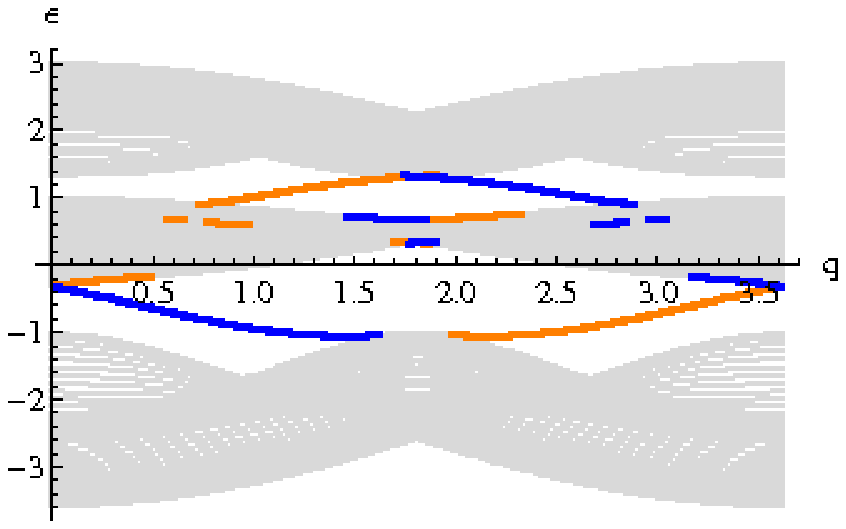}
			\label{fig:subfig6}
		}			
	\caption{Spectrum of the modified Kane-Mele model in Eq.~\eqref{H_2}, with a
	metallic (triangular) band added
	(Fig. \ref{fig:model_conventions}). Here we use the parameters $t_1 = 1, \lambda = 0.5, t_2 = 0.2, h = 0.4$, and vary $t_3$.
	We show spectra for both zigzag (\ref{fig:subfig1}, \ref{fig:subfig3}, \ref{fig:subfig5}) and armchair edge strips
	(\ref{fig:subfig2}, \ref{fig:subfig4}, \ref{fig:subfig6}), with mixing parameter values $t_3 = 0,0.03,0.3$.
	The strips are finite in one direction, and periodic in the other, so the lattice is wrapped around a cylinder.
	The edge-state branches on different sides of the sample are denoted by thick lines, dark (blue) and light (orange) respectively. 
	All bulk states are colored light gray. Helical edge state branches can still appear (at least when the Fermi energy is shifted),
	as can be easily seen by counting the (odd) number of surface state branches on one side of the strip 
	crossing $E=0$ between $q=0$ and the BZ midpoint ($q = \pi$ for zigzag, $q=\pi/\sqrt{3}$ for armchair).
	}
	\label{fig:3band_numerics}
\end{figure}

Our starting point is the first 2D model of a topological insulator, 
the Kane-Mele model\cite{Kane:2005A,Kane:2005B}, defined on a honeycomb lattice. The first construction proposed above can be realized by
considering this model coupled to a very simple 
metallic band: we add lattice sites at the centers of the honeycomb plaquettes, as shown in Fig.~\ref{fig:model_conventions}, so that the 
new sites are now those of a triangular lattice. To form a metallic band we allow hopping between the nearest-neighboring new (triangular) sites. 
With a Fermi energy crossing anywhere in this band, we have a single Fermi surface, centered about ${\bf q}=0$. We then couple between 
the two parts of our model, by allowing hopping between the honeycomb lattice sites and the new triangular lattice sites
(see Fig.~\ref{fig:model_conventions}). The Hamiltonian is
\be\label{H_2}
\begin{split}
{\mathcal H} = &
- t_1 \sum_{\langle i j \rangle \alpha} \left[a^{\dagger}_{i \alpha} b^{\phantom\dagger}_{j \alpha} + h.c. \right]
- t_2 \sum_{\langle i j \rangle \alpha} \left[ c^{\dagger}_{i \alpha} c^{\phantom\dagger}_{j \alpha} + h.c. \right]
+ i \lambda \sum_{\langle \langle i j \rangle \rangle \alpha \beta} 
\left[
a^{\dagger}_{i \alpha} a^{\phantom\dagger}_{j \beta} s^z_{\alpha \beta} \nu_{i j}
+ h.c. + \left( a \rightarrow b \right)  \right]
\\ &
- t_3 \sum_{\langle i j \rangle \alpha} \left[ c^{\dagger}_{i \alpha} a^{\phantom\dagger}_{j \alpha} +
c^{\dagger}_{i \alpha} b^{\phantom\dagger}_{j \alpha} + h.c. \right]
+ h \sum_j c^{\dagger}_j c^{\phantom\dagger}_j
\; ,
\end{split}
\ee
where $i,j$ denote all the combined lattice sites, $a,b,c$ denote the fermion operators on the three sublattices ($a,b$ 
for the honeycomb, and $c$ for the triangular metal, see Fig.~\ref{fig:model_conventions}). 
In addition, spin indices are denoted by $\alpha,\beta$, and $\nu_{i j}$ is as defined in Ref~\onlinecite{Kane:2005A,Kane:2005B} 
(and shown in Fig.~\ref{fig:model_conventions}).
For convenience we have included an independent chemical potential $h$ for the triangular lattice.

Following closely the procedure in Refs.~\onlinecite{Kane:2005A,Kane:2005B},
we calculate the spectrum of Eq. (\ref{H_2}) using exact diagonalization of strips that
are terminated either at a zigzag or an armchair edge, as these are
representative edge cuts of generic boundaries of the honeycomb lattice.
We calculate for one spin polarization in order to avoid clutter, as $s^z$ is a good
quantum number in our model, and deviations from this are immaterial to the physics we explore here,
and identify explicitly the eigenstates localized at each edge of the finite strip. 
We present the spectrum of the model in Fig.~\ref{fig:3band_numerics}, when the honeycomb-triangular hybridization parameter
is zero, weak and strong ($t_3 = 0, 0.03, 0.3$ respectively),
while keeping the other model parameters fixed at $t_1 = 1, \lambda = 0.5, t_2 = 0.2, h = 0.4$. 
We choose the Bravais lattice vectors to be 
${\bf a}_1 = a {\hat x},{\bf a}_2 = \frac{a}{2} \left( - {\hat x} + \sqrt{3} {\hat y} \right) $. 
The strip widths we use are 30 unit cells, and are terminated in a symmetric fashion. For the zigzag geometry, states 
identified as edge states on one side of the system have at least $0.85$ of their total weight
within a distance of $y = 30 \frac{a}{4}$ from the lower edge (at $y=0$), and the edge states at the other 
side of the sample have at least $0.85$ of their weight fraction above $y = 30 \frac{3 \sqrt{3} a}{4}$ 
(for the armchair edge strip, we take the limits for this procedure to be 
$x<30 \frac{a}{4 \sqrt{3}}$ and $x> 30\frac{3 a}{4}$, and consider $0.7$ 
of the total wavefunction weight rather than $0.85$).

\paragraph{Weak hybridization.} In Figs.~\ref{fig:subfig3},
\ref{fig:subfig4}, we plot the spectrum of bulk and edge state for an
armchair and zig-zag strip with weak hybridization, $t_3=0.03$.
Comparing with the decoupled spectrum in Figs.~\ref{fig:subfig1},
\ref{fig:subfig2}, the numerical results clearly show that essentially the same helical surface states appear
in our model without a bulk gap present - the helical surface states are only weakly perturbed.
Only in those areas where the helical surface states overlap in both energy and momentum with bulk states, 
do we see any appreciable change.
Since different momenta cannot mix in the absence of disorder, it is not surprising the surface states are unaffected
if they do not overlap in both energy and momentum.

\paragraph{Strong hybridization.} We consider also {\it strong} coupling ($t_3=0.3$) between
the honeycomb and triangular subsystems. The spectra, shown in Figs.~\ref{fig:subfig5},\ref{fig:subfig6},
in which, surprisingly, the surface state branches 
have been pushed out of those regions where the bulk states reside,
away from their original position, toward those 
regions of the momentum-energy diagram where bulk states are
absent. This results in the doubling of the number of surface state
branches for a single momentum, as can be seen in
Fig. \ref{fig:subfig5} in the range $2.4<q<3.2$. 
One branch spans the region below the metallic density
of states, and another above.

It is noteworthy that for other parameter choices in this model, one
can also find instances where the number of co-moving surface states
changes upon bulk-edge coupling, but still with an odd number of surface
branches on each side of the strip (in the supplementary material,
Section~\ref{Three_states}, an example is given where the number of
surface states changes from 1 to 3 - still retaining the odd number of branches). 

Our numerical results suggest a tendency of surface states to be pushed away 
from overlapping in (surface-projected) momentum and energy with
parasitic bulk states originating from a non-topological band. Not
only do they seem to persist, in some momentum number they seem to
multiply. Motivated by this observation, we turn 
next to an analytic treatment of the mixing between surface states and bulk states.


\section{The Fano model applied to edge-states - bulk mixing}
\label{Fano}
 
Our numerical results confirm that edge states have remarkable
resilience. Even in the absence of a band gap in the material, precise
edge states form in empty spaces in the projected energy-momentum
state diagrams. Let us now approach the problem in its idealized form
analytically. For this purpose we write a model consisting of a surface state branch and a decoupled bulk metallic band, and then allow them to mix.
This most certainly describes the specific model we have discussed here, since an effective 
low-energy continuum theory would encompass precisely these elements.
\footnote{For concreteness, we consider a single surface state branch running along a domain wall in an 
otherwise bulk system. This is produced by a soliton mass in the appropriate continuum Dirac model 
for the topological insulator bulk, the surface states being the
localized solutions first found in Ref.~\onlinecite{Jackiw:1976}}.
We label the transverse momentum by ${\bf q}$, and the momentum in the direction 
perpendicular to the wall by $k$. In the clean limit, the bulk
and surface states are mixed by hopping matrix elements which preserve
${\bf q}$, and therefore states with different 
${\bf q}$ do not mix. This allows us to treat individual edge states
separately. 

The model outlined above takes the form of the well known Fano
model\cite{Fano:1961}. Making use of the path-integral formulation, the action for the model is 
\be\label{Fano_1}
\begin{split}
S_1 & =
\frac{1}{\beta} \sum_n \int_{\bf q}
\Bigg[
\psi^{\dagger} \psi \left[i \omega_n - \epsilon({\bf q}) \right]
+
\int_{k} \chi^{\dagger} \chi \left[ i \omega_n - E(k,{\bf q}) \right]
+
\int_{k} \left[ \psi^{\dagger} \chi \, g(k,{\bf q}) + h.c. \right]
\Bigg]
\; ,
\end{split}
\ee
where the fermionic Matsubara frequencies are $\omega_n = \pi \frac{2n+1}{\beta}$,
and the surface and bulk states are denoted by the Grassmann fields $\psi = \psi(i \omega_n,{\bf q})$, 
and $\chi = \chi(i \omega_n,k,{\bf q})$ respectively, their dependence on momentum and Matsubara frequency
suppressed for the sake of brevity.
The surface and bulk state energies are $\epsilon({\bf q})$ and $E(k,{\bf q})$ respectively,
and the coupling between them is $g(k,{\bf q})$. The exact solution of the Fano model\cite{Mahan}, 
can be most easily achieved by integrating out the bulk degrees of freedom. Since the action is quadratic, 
this can be done exactly, resulting in an effective action for the 
surface state degrees of freedom ($\psi$) alone:
\be\label{Fano_2}
S_2 =
\frac{1}{\beta} \sum_n \int_{\bf q}
\psi^{\dagger} \psi \left[ i \omega_n - \epsilon({\bf q})
- \int_k \frac{ |g(k,{\bf q})|^2 }{i \omega_n - E(k,{\bf q})}
\right]
\; .
\ee
This action reveals the fate of the spectrum of the surface states. The retarded Green's function extracted from the effective action above is
\be\label{Green_ret}
G_{ret}^{-1} = \left[ \omega - \epsilon({\bf q})
- \int_k \frac{ |g(k,{\bf q})|^2 }{\omega - E(k,{\bf q}) + i \delta}
\right]
\; ,
\ee
$ $From which the spectral function $A({\bf q},\omega) = - 2 Im\left[G_{ret}({\bf q},\omega)\right]$ 
can be extracted.

\begin{figure}
	\centering
		\includegraphics[width=3.0in]{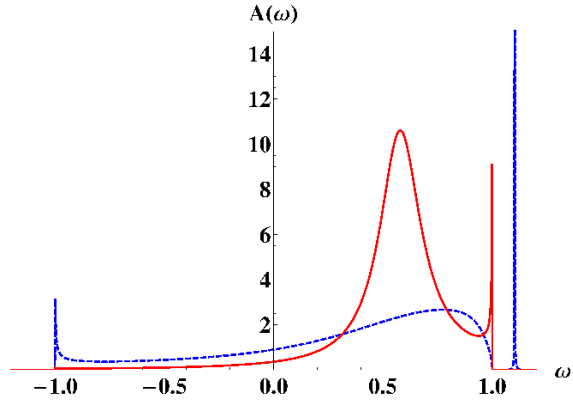}
	\caption{Typical spectral weight function of the Fano model. We take the spectral weight in Eq~\eqref{Spectral_func},
	with parameters $E_1 = -1$, $E_2 = +1$, the surface state energy $\epsilon = 0.5$
	(within the bandwidth). For $\omega$ values outside the bulk bandwidth we take $\lambda = 10^{-4}$ to approximate the delta function outside
	the bandwidth. For the continuous (red) curve we take $|g|^2 \nu(E_F) = 0.04$, and for the dashed (blue) curve we take $|g|^2 \nu(E_F) = 0.2$.
	At the lower coupling value (continuous curve), the widened peak near $\omega = 0.5$ is the remnant of the original surface state, and the modest
	weight in one delta function already appears very near the band edge (the weight of the other delta function is too small to discern). 
	For the stronger coupling value (dashed line) traces of the surface state
	are largely obscured in the bulk bandwidth, but both the delta-functions outside the bandwidth appear prominently. 
	}
	\label{fig:Fano_spectral_weight}
\end{figure}

The most important feature of the spectral function $A({\bf
  q},\omega)$ is that the original surface states, when their energy
is within the bandwidth of the bulk, acquire a lifetime, which is roughly 
$\tau \sim \frac{1}{|g|^2 \nu(E_F)}$ (where $\nu(E)$ is the density of
  states, and $E_F$ the Fermi energy). This lifetime describes the
  typical time scale in which the edge probability density leaks into the
  bulk states. In addition, new delta functions appear at energies \emph{outside}
the bandwidth of the bulk metallic band both above and below. We illustrate this by considering a constant coupling $g$,
and a uniform density of states $\nu(E_F)$, in a bulk band with energies $E_1 < E < E_2$. The spectral function is
\be\label{Spectral_func}
A(\omega) = 
\frac{- 2 Im(\Sigma) }{\left[ \omega - \epsilon - \frac{\lambda}{2} \log \left(\frac{(\omega - E_1)^2}{(E_2-\omega )^2}\right) \right]^2 + Im(\Sigma)^2}
\; ,
\ee
where $\lambda = |g|^2 \nu(E_F)$, and $Im(\Sigma) = -\pi \lambda$  for
$E_1 < \omega < E_2$, and \emph{zero} if $\omega$ lies outside the
bandwidth of the bulk band, thus producing the aforementioned delta
functions. Fig.~\ref{fig:Fano_spectral_weight} shows the spectral function for particular 
values of the parameters. If the original surface state energy lies outside the bulk bandwidth,
it remains as a delta function outside the bandwidth, slightly shifted from the original surface state energy, and with most of its spectral weight retained. In addition, a new delta-function peak will appear on the other side (in energy) of the bulk band, with a small spectral weight, and some spectral weight will appear within the bulk bandwidth.

These features of the Fano model depend only weakly on the details of the density of states,
as can be demonstrated by considering a more realistic bulk band dispersion $E(k) = -\mu - 2t \cos(k)$. 
Using the dimensionless variable $u \equiv (\omega + \mu)/2t$, 
the integral for the self-energy can be solved analytically, yielding 
$Re(\Sigma) = \frac{g^2}{t} \frac{sgn(u)}{\sqrt{u^2-1}}$ and $Im(\Sigma) = 0$
when $|u| >1$, and $Re(\Sigma) = 0$ and $Im(\Sigma) = -\frac{g^2}{t\sqrt{1-u^2}}$ when $|u|<1$.
The resulting spectral function is qualitatively no different than
that of Fig.~\ref{fig:Fano_spectral_weight}, which corresponds to Eq. (\ref{Spectral_func}).

The remarkable structure the Fano model implies can be further intuitively understood by considering an impurity state ($d$) coupled to a flat band of bulk states 
($f_n$) at zero energy,  
${\mathcal H} = \epsilon d^{\dagger} d + g \sum_n \left[ f_n^{\dagger}
  d + h.c. \right]$ (the coupling $g$ is taken constant without loss
of generality), with $n = 1 \ldots N$. The characteristic polynomial of the 
Hamiltonian matrix $Det(E - {\mathcal H}) = (-1)^{N-1} E^{N-1} \left( E^2 - \epsilon E - N g^2 \right)$,
has $N-1$ zero modes left from the original $N$ flat band states, and two roots at
$E = \epsilon/2 \pm \sqrt{\epsilon^2/4 + N g^2}$, outside the flat band,
which, assuming $N g^2\gg \epsilon^2$, become $E \approx \pm \sqrt{N} g + \epsilon/2$.
As in the Fano model above, exact energy eigenstates appear above and below the band.
While this is an extremely artificial example (flat band states being most sensitive to coupling to other states), 
it demonstrates how the salient features of the Fano model can be understood - 
the impurity state ``hijacks'' one effective mode from the band, and mixes with it, producing two eigenstates with energies 
outside the band, which correspond to reduced-weight delta functions in the spectral function for ${\hat d}$.

\subsection{Fano model implications}
\label{Fano_B}

We can interpret the numerical results of Section~\ref{numerics} in terms of the salient features of the Fano model as follows.
The Fano model tells us that for those edge-parallel momenta ${\bf q}$
where the surface states and bulk states \emph{do not} overlap in energy, the surface states are only slightly shifted in energy.
This is demonstrated in those regions in Fig.~\ref{fig:3band_numerics} where the original surface states did not overlap with the bulk states.
On the other hand, if the surface states overlap in energy with the bulk 
state energies, the surface states are ``exiled'' from the bulk bandwidth, and form states at (very different) lower and higher energies,
in addition to leaving ``ghost'' surface resonances overlapping with the bulk states,
close to where they originated.
The ``exiled'' states are still surface states, mixed with a superposition of bulk states localized at the surface,
since any eigenstate existing in a bulk gap in energy-momentum space, must be an evanescent wave into the bulk.
The surface state spectrum $\epsilon({\bf q})$ is a continuous function of ${\bf q}$,
and assuming the coupling $g(k,{\bf q})$ is also a continuous function of momentum,
the exiled surface states will also form a ${\bf q}$ - continuous energy branch.
The helical nature of the surface states will also be preserved in the new exiled surface states: 
it will correspond to an odd number of Dirac cones in 3D (and an odd number of Kramer's pairs of bands in 2D).
The exiled states are evident in the numerics in Fig.~\ref{fig:subfig5} and Fig.~\ref{fig:subfig6},
in particular in Fig.~\ref{fig:subfig5} in the range $2.4 < q < 3.2$.
Indeed, the exiled states retain the surface state branch continuity, as well as their helical nature.
The ghost surface resonances are not pure surface states, and so are not distinguished in the spectra in Fig.~\ref{fig:3band_numerics}.
In order to identify the ``ghost state'' signatures in our numerical, we probe the explicit spectral weight 
function of our model in the next subsection.

Experimentally, we expect that on different facets of the crystal the two 
scenarios could be realized, and so if a helical surface state appears on one facet of the metal, crossing the Fermi energy, but not 
on the other facets, it may simply be significantly shifted in energy, and can in principle still be observed outside the energy range of 
the bulk band.


\subsection{Numerical evidence for the surface resonance}
\label{spectral_weight}

\begin{figure}
	\centering			
	\includegraphics[width=3.0in]{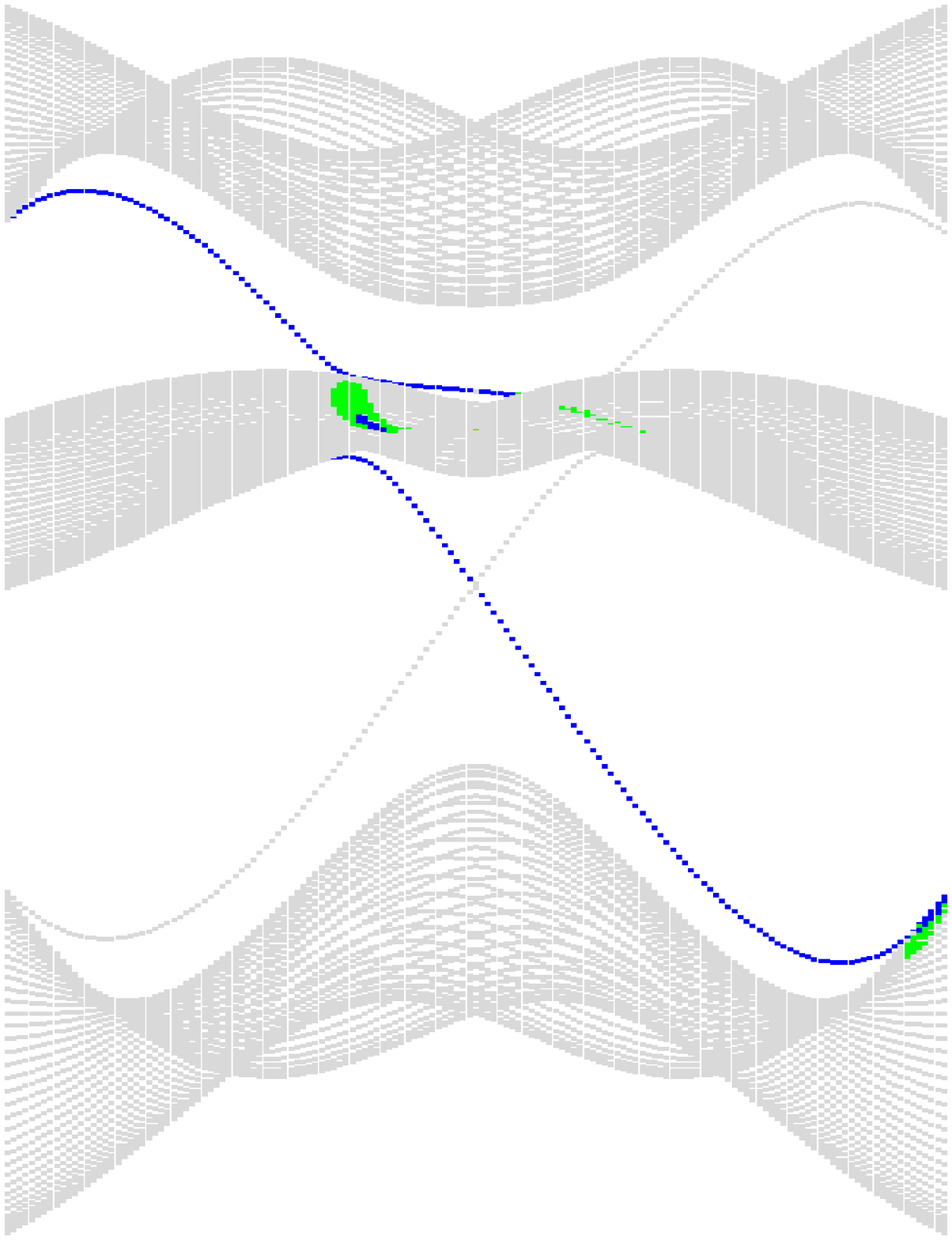}
	\caption{Numerical calculation of the surface state spectral function. We take the surface state eigenstates in the decoupled case ($\psi_0({\bf q})$)
	and find their distribution among the eigenstates of the coupled system $\phi_{\omega}({\bf q})$ by calculating the overlaps
	$A({\bf q},\omega) = |\dirac{\psi_0({\bf q})}{\phi_{\omega}({\bf q})}|^2$. This is expected to approximate Eq.~\eqref{Spectral_func}. 
	We use the same parameters as the strongly hybridized zigzag edge strip in Fig.~\ref{fig:subfig5}, and the overall spectrum is manifestly identical 	
	in the two images. We plot the surface state spectral weight function versus transverse momentum (horizontal axis)
	and energy (vertical axis). Dark (blue) points have $A(\omega)>0.06$, very light (green) points have $0.06>A(\omega)>0.02$,
	and all points with $0.02>A(\omega)$ are gray.
	The exiled surface state branches are manifest, and the ghost spectral resonance is identifiable as a faint
	peak overlapping with the bulk band, between the two exile branches. In addition, the lower spectral weight points map out the shadow of the 
	full bulk spectrum.
	}
	\label{fig:Spectral_Function_numerical_image}
\end{figure}

In this section we will present numerical evidence for the ``ghost state'' surface resonance in the model described in section~\ref{numerics}.
First, we calculate explicitly the surface state spectral weight function \eqref{Spectral_func} described above in the Fano model analysis. 
This is done by calculating a probability distribution to find the original surface state eigenstates at a given energy and momentum in the new, coupled spectrum.
We take the surface state eigenstates $\psi_0({\bf q})$ in the decoupled case, and calculate
their overlap with the various eigenstates of the coupled system, squared.
In particular, if we denote by $\phi_{\omega}({\bf q})$ all the eigenstates of the coupled system, with energies $\omega$, 
the spectral weight is $A({\bf q},\omega) = |\dirac{\psi_0({\bf q})}{\phi_{\omega}({\bf q})}|^2$.
The results of our numerics, for the surface state branch along one edge of the sample in the strong coupling case ($t_3 = 0.3$), are shown in Figs.~\ref{fig:Spectral_Function_numerical_image}. A faint but discernible diffuse peak is seen to overlap with the bulk states in between
the upper and lower exiled surface state lines, thus confirming the predictions from the Fano model analysis.

\begin{figure}
	\centering			
		\subfigure[ ]{
			\includegraphics[width=3.0in]{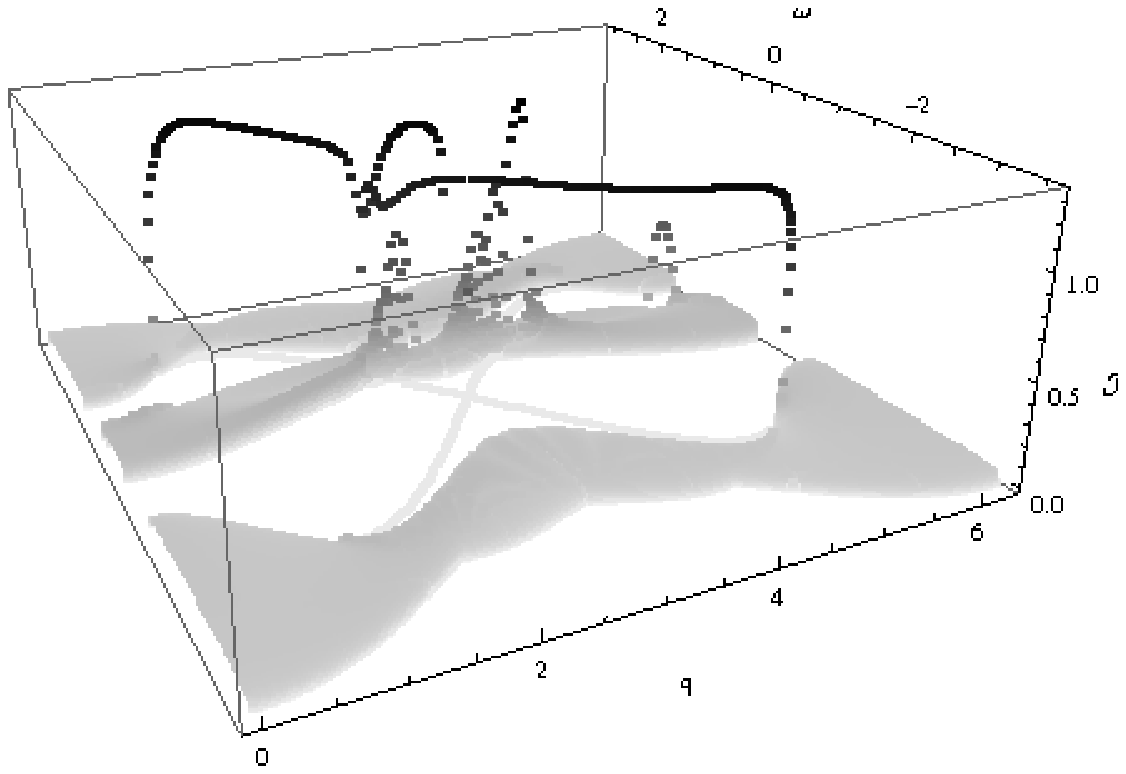}
			\label{fig:subfigD}
		}
		\subfigure[ ]{
			\includegraphics[width=3.0in]{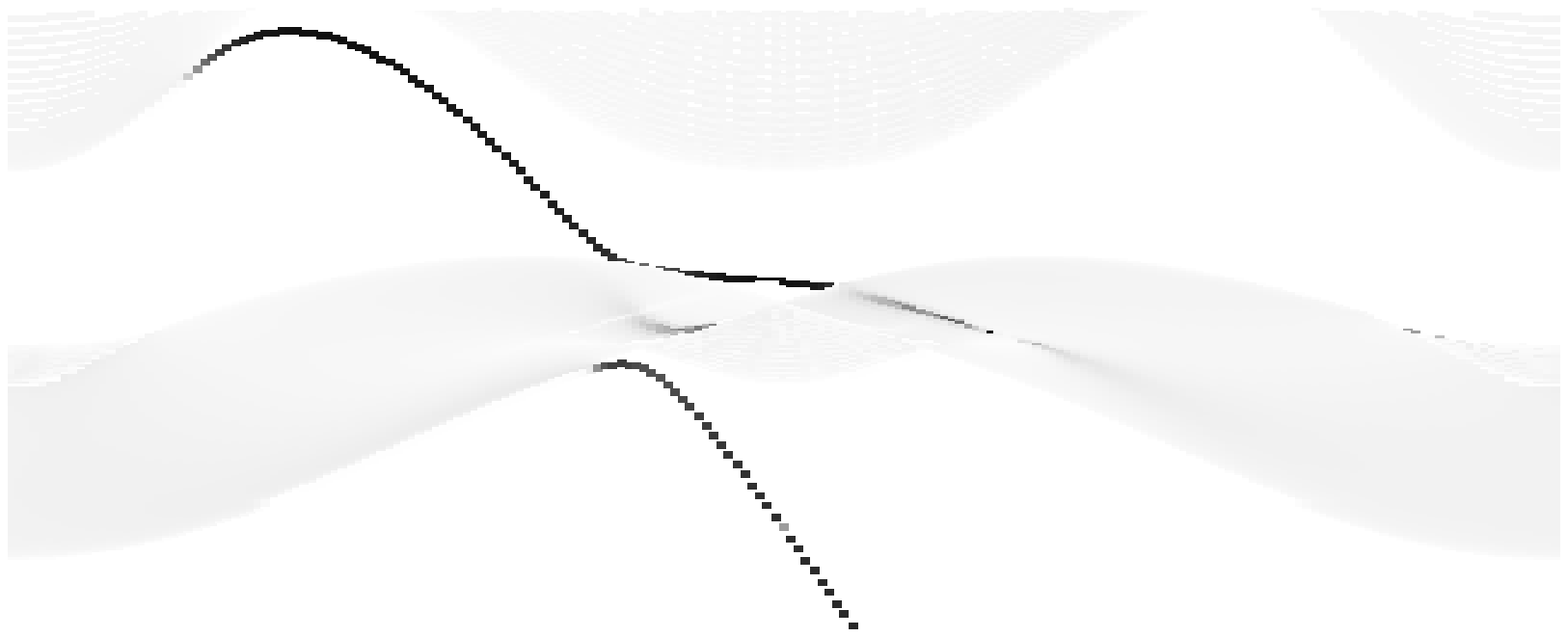}
			\label{fig:subfigE}
		}		
	\caption{The `ghost' surface resonance observed in surface probes. The quantity ${\mathcal J}$ measures
  the weight of each eigenstate $\phi_{\omega}$ convolved with an exponentially decaying weight factor,
  mimicking the penetration depth of the experimental surface probes
  ${\mathcal J}(\omega,{\bf q}) = |\bra{\phi_{\omega}({\bf q})} f(y) \ket{\phi_{\omega}({\bf q})}|^2$.
	Here we plot ${\mathcal J}$ versus transverse momentum ${\bf q}$ and energy $\omega$,
	both in a 3D (a) and 2D (b) plot. The gray-scale color scheme is such that higher values are colored 
	dark gray. We find significant values of ${\mathcal J}$ only for the surface states and the surface resonances, which are 
	manifestly positioned in between the exiled surface states (most clearly in the 2D plot (b).
	In addition, the very light gray points map out the bulk states.
	}
	\label{fig:surface_resonances}
\end{figure}

The spectral weight function \eqref{Spectral_func}, is not necessarily what ARPES or STM will measure.
In order to give a clear experimental signature that can be measured, we perform one additional numerical analysis.
We take the density profile of each eigenstate $\phi_{\omega}({\bf q})$, and convolve it with a weight factor $f(y) = e^{-\frac{y}{4a}}$,
to yield the total weight ${\mathcal J}(\omega,{\bf q}) = |\bra{\phi_{\omega}({\bf q})} f(y) \ket{\phi_{\omega}({\bf q})}|^2$. 
The decaying exponential mimics the finite penetration depth surface probes can achieve. 
Advantageously, the calculation of ${\mathcal J}$, is unbiased by the ``band archeology'' in calculating $A({\bf q},\omega)$,
which required comparison with the eigenstates of the decoupled system. 
We plot ${\mathcal J}(\omega,{bf q})$ versus $\omega$ and ${\bf q}$ in Fig.~\ref{fig:surface_resonances}. 
Those eigenstates with a significant part localized at the surface should
have a sizable value of ${\mathcal J}$. Indeed, we see in Fig.~\ref{fig:surface_resonances} that between the exiled surface states the 
signature of the ghost surface resonance appears. In conclusion, careful analysis of our numerical results 
shows that surface state probes could identify the ``ghost'' surface resonance.

\section{Disordered helical metal within the Fano model\label{disorder}}

Going beyond the clean limit, we now consider edge-bulk hybridization
in the presence of disorder. A random disorder potential scatters
surface states into other surface states, as well as into the
bulk. For simplicity we
ignore, however, disorder scattering between bulk states,
concentrating only on the fate of the surface states. Within the
weak-disorder approximation \cite{Edwards:1958} we
will see that the Fano picture by and large still applies.

Disorder scattering can easily be quantified by slightly modifying
Eq. \eqref{Fano_1}:
\be\label{Disorder_1}
\begin{split}
S_3 & =
\frac{1}{\beta} \sum_n \int_{\bf q}
\Bigg[
\psi^{\dagger} \psi \left[i \omega_n - \epsilon({\bf q}) \right]
+
\int_{k} \chi^{\dagger} \chi \left[ i \omega_n - E(k,{\bf q}) \right]
+
\int_{k,{\bf q}'} \left[ \chi^{\dagger} \psi' \, g(k,{\bf q,q'})^* + h.c. \right]
+
\int_{\bf q'} \psi^{\dagger} \psi' V({\bf q,q'})
\Bigg]
\; ,
\end{split}
\ee
where we use the shorthand notation $\psi' \equiv \psi(i \omega_n, {\bf q'})$ (and similarly for $\psi''$ later on).
The coupling term $g(k,{\bf q,q'})$ no longer conserves transverse momentum,
and is determined by the overlap of the surface (${\tilde \psi}$) and bulk (${\tilde \chi}$) state
wavefunctions represented by the various fields weighted by the random potential,
$g(k,{\bf q,q'}) = \int_{\bf r} {\tilde \psi}_{\bf q'}({\bf r})^* U({\bf r}) {\tilde \chi}_{k,\bf q}({\bf r})$.
Integrating out the bulk fermions, we are left with the effective action:
\be\label{Disorder_2}
\begin{split}
S_4 & =
\frac{1}{\beta} \sum_n
\Bigg[
\int_{\bf q} \psi^{\dagger} \psi  \left[i \omega_n - \epsilon({\bf q}) \right]
+
\int_{\bf q,q'} \psi^{\dagger}_{\bf q} \psi^{\phantom\dagger}_{\bf q'} V({\bf q,q'})
-
\int_{k, {\bf q,q',q''}} 
(\psi')^{\dagger} \psi'' 
\frac{  g(k,{\bf q,q'})^* g(k,{\bf q,q''})  }{i \omega_n - E(k,{\bf q})}
\Bigg]
\; ,
\end{split}
\ee

Next we average over disorder, assuming the disorder has a Gaussian distribution
$\overline{U({\bf r}) U({\bf r}') } = C({\bf r - r'}) = {\tilde u}^2 \delta({\bf r - r'})$, 
and surface and bulk wavefunctions
${\tilde \psi} \sim e^{i {\bf q \cdot R} - |x|/d}$ and ${\tilde \chi} \sim e^{i {\bf q \cdot R} + i k x}$ 
(where the edge-parallel and perpendicular coordinates have been separated as ${\bf r} = (x,{\bf R})$,
and $d$ is the skin depth of the surface states).
Finally, expanding in the limit of weak disorder we find the leading contribution from disorder is a 
term \emph{quadratic} in $\psi$ (higher order terms in $\psi$ are weaker in this limit)
\be\label{Disorder_3}
\begin{split}
S_5 & = 
\frac{1}{\beta} \sum_n
\int_{\bf q} \psi^{\dagger} \psi 
\Bigg[
\left[i \omega_n - \epsilon({\bf q}) \right]
- \int_{k, {\bf q'}}
\frac{u^2}{i \omega_n - E(k,{\bf q'})}
\Bigg]
\; ,
\end{split}
\ee
where $u$ differs from ${\tilde u}$ by some numerical constant. 
It is important to note that this leading term originates solely from the bulk-surface state scattering,
and can also be derived by simply taking the disorder mean of the action in \eqref{Disorder_2}. This gives a 
Green's function nearly identical to that of \eqref{Green_ret}, but we notice that while transverse momentum is now once again a good quantum number,
disorder, even after averaging, couples \emph{all} bulk states to any one of the surface states. Therefore, the separation in energy and momentum picture no longer holds here, and any surface state that does not exist in a full energy band gap will suffer the effects of a Fano mode impurity 
state in the energy bandwidth of the bulk band. From the original surface state branch, a ``ghost'' feature will remain in the spectral function, and
``exiled'' surface state delta-functions appear outside the bulk bandwidth. When disorder is very weak, the ``ghost'' surface 
states will actually be the most noticeable feature, and will simply seem as ordinary surface states with a lifetime to leak into the bulk, 
determined by the strength of the disorder.

A simple experimental test for our predictions follows from the observation
that the scattering strength is proportional to impurity density: $u^2\propto \rho_{imp}$. In samples where the disorder scattering of 
surface states into the bulk dominates over other effects (phonon
scattering and electron-electron interactions), the surface-state
lifetime must be inversely proportional to the impurity density.  Very
recent work\cite{Park:2010} has already measured a quasiparticle
lifetime for the surface states in Bi$_2$Se$_3$, which is one of the
main candidates to be a helical metal as mentioned above, and
concluded from the energy dispersion of the surface state lifetime
that disorder scattering seems to be the dominating scattering
mechanism. It would be particularly interesting in such experiments to
search for the exiled states, which should be above and below the bandwidth of the bulk band.
We leave a theoretical study of the nature of
the exiled states in disordered helical metals to future work.
 
The difference between helical surface states and non-helical surface
states becomes paramount when considering disorder.  Non-helical surface states, in the clean limit, consist of an even number of bands that can mix via time-reversal 
preserving terms.  In 2D materials with 1D surface states, non-magnetic disorder, of the likes we consider here, can back-scatter between two 
such bands since their spin configurations are non-orthogonal, and therefore
strongly localize them (although a strong spin-orbit interaction may
mitigate this effect \cite{Xu:2006}). A helical edge state on the other hand is a
chiral 1D conductor, which suffers no backscattering. In 3D 
materials the surface states are 2-dimensional, and non-magnetic
disorder would lead to weak localization\cite{ALRA:1979}. 3D
helical surface states, however, always exhibit anti-localization of a single Dirac cone. 
Non-helical surface states may also be in an anti-localization class due to spin-orbit coupling, but unlike the helical states, 
they are (topologically) smoothly connected to a spin-rotationally symmetric 2DEG, 
which suffers weak localization. The additional protection that
helical surface states exhibit against localization effects indicates
that the surface-bulk Fano effects will dominate over localization
effects, whereas the opposite may be true for non-helical systems.

\section{Conclusions}

Topological band insulators are characterized by their surface state properties.
Helical states on the surface of a TI have \emph{odd}, rather than even, number of either Kramer's pairs of 1d
surface channels in a 2d material, or 2d surface metals with a Dirac
dispersion in a 3d
material\cite{Kane:2005A,Kane:2005B,Bernevig:2006,Moore:2007,Fu:2007}.
In our work we demonstrated that even materials that are generically metals
retain features of the helical structure of the surface states, and therefore a similar classification can apply to metallic states.

Furthermore, we find that the mixing between surface states and bulk
states in metals results in a generic rearrangement of the
surface-state spectrum. At energies where surface and bulk states
overlap and mix (either when they overlap also in momentum parallel to
the surface, or due to disorder)  the surface states are reduced into
surface resonance ``ghosts'', with a diffused spectral function peak centered
near the original energy of the surface states. In addition, the
bulk-surface mixing produces ``exiled surface states'' outside the
confines of the bulk parasitic metal energy bandwidth.
Therefore, the surface-state spectral structure emerging from this mixing, 
contains for each surface momentum ${\bf q}$ (where surface
states exist) two sharp surface states above and below the parasitic metal bands
where there is a bulk gap (for that momentum number, or, when weak disorder
is at play, a complete bulk gap), and a diffuse surface resonance
overlapping with the metallic density of states. 
Surface resonances, despite being immersed in the bulk states, can still be identified in ARPES measurements due to the fact that they have 
very weak $k_{\perp}$-dispersion\cite{Kevan:prb1991} (and ideally none), in contrast to bulk states.
Following the surface resonance peak for different ${\bf q}$'s should roughly parallel
the sharp surface states spectrum above and below the band. 
The location of the sharp exiled states and the diffuse ghost resonance
may change for surfaces made of different facets (cuts) of the
crystal. For instance, at a particular energy a sharp edge state on
one facet may correspond to a mere ghost resonance when we consider another
surface. 

The evolution of the surface-state spectrum in the hybridized system, as described above, 
makes it clear that that the helical characteristic of the surface states
will remain unchanged. Strikingly, in some cases the number of surface state branches at each edge can change, but
the parity of the number of surface-bands will remain odd (see supplementary material Section~\ref{Three_states},
where the 1 surface state branch gets multiplied to 3).

Effects of disorder were only briefly and crudely considered here. Nevertheless we can already indicate
intriguing features which may arise. Following the disordered Fano model for the
case of a metallic band overlapping in energy but not in momentum, the
hybridization will broaden a surface state (say at momentum ${\bf q}$)
into a ghost, but will also produce sharp (surface) energy-eigenstates
at energies above and below the energy overlap range, so long as
there is an empty patch as a function of energy that can support
them. An ARPES measurement should be able to observe all these
features for metals with an appropriate band structure, where a
metallic parasitic band overlaps in one range of energies with
topological edge states, but not in momentum, and above or below this
energy range bulk gaps exist. Tunneling measurements should also be able to observe the
spectral features we describe here. We will explore this situation more
closely in future work. A more mundane prediction of the disorder analysis which should be
easily accessible in experiment is the quasiparticle lifetime
decrease due to increasing disorder.

Topological insulators (and topological phases in general) are characterized by a topological invariant
that assumes only quantized values, and is thus robust to infinitesimal deformations to the model.
The presence of edge states is deeply connected to topological order, and the presence of helical edge
states in the metallic models we present here would suggest some sort of topological order may exist.
However, conventional understanding of topological phases depends crucially on having a robust global gap in the system.
The `helical metal' phase we consider here is gapless, though in a clean system local gaps in the BZ may appear.
It is therefore unclear whether one can define a topological invariant in general. However, one incarnation of the topological invariant
for the topological insulators (in both $d=2,3$), involves Bloch states only at time-reversal invariant momenta in the BZ\cite{Fu:2007}.
If there is a local gap at these points in the BZ, the same topological invariants are still well defined.
In the `helical metal' phases we introduce here, each of these points can either be locally gapped, or not.
It is also unclear what could be a topological invariant for the `helical metal' in the presence of disorder.
Going forward, the question whether a topological invariant exists for the systems we introduce here is perhaps the most enticing,
and we leave its determination to future work.

The abundance of materials exhibiting topological properties suggests
that there must be many materials which are  'helical metals' - metals with an odd
number of chiral surface states at some energy ranges for each facet. Such materials can presumably be found in the
vicinity of topological insulators and vice versa, which is supported
by the materials observed so far, Sb, Bi$_{1-x}$Sb$_x$, and Bi$_2$ Se$_3$. 
The spectral effects which we explore here should be accessible in all of these materials and
provides another challenging system where interaction and disorder may
have important and interesting effects. 

It is a pleasure to acknowledge useful discussions with P. A.
Lee, O. Motrunich, Z. Hasan, and D. Hsieh.
This work was supported by the 
Sherman Fairchild Foundation, by the Packard foundation, Sloan
fellowship, and Cottrell fellowship, and by the Institute for Quantum Information under NSF
grants PHY-0456720 and PHY-0803371. 

\appendix

\section{Supplementary material}
\label{supp}

\subsection{Alternative construction of a helical metal model}
\label{2band_model}

\begin{figure}
	\centering
		\subfigure[ Zigzag edge]{
			\includegraphics[width=3.0in]{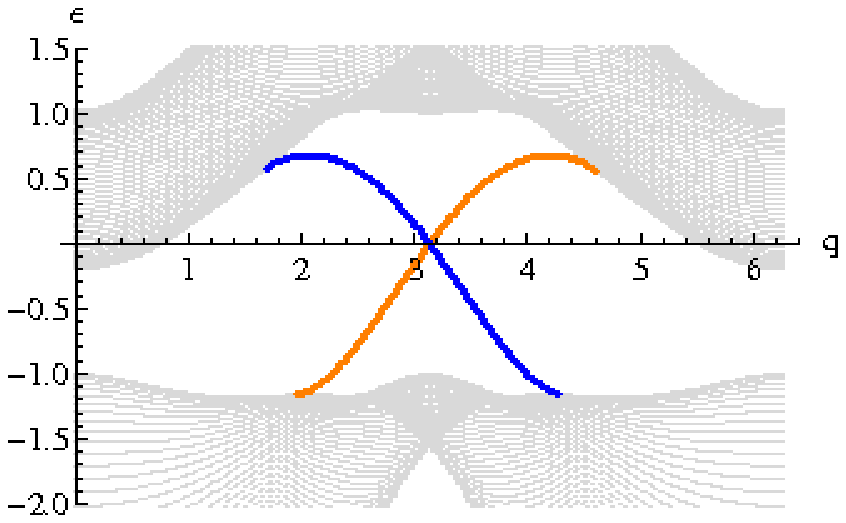}
			\label{fig:subfig12}
		}			
		\subfigure[ Armchair edge]{
			\includegraphics[width=3.0in]{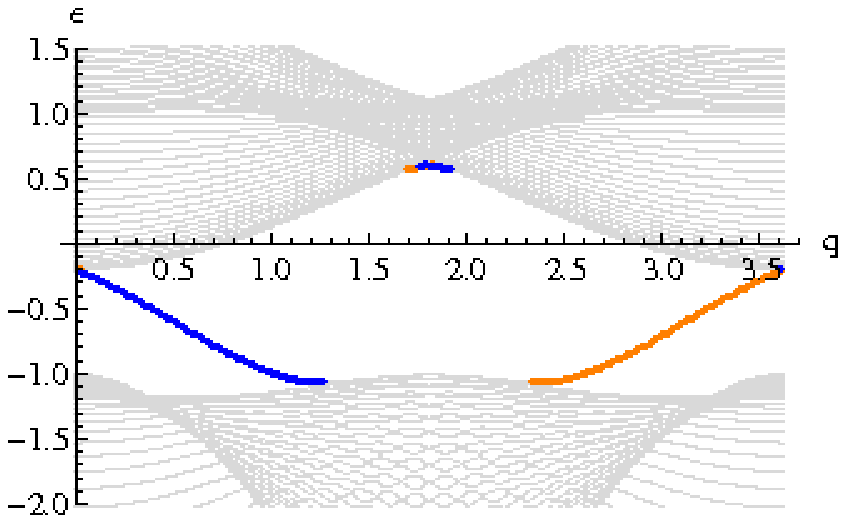}
			\label{fig:subfig13}
		}			
	\caption{
	All pictorial conventions are exactly the same as in Fig.~\ref{fig:3band_numerics}, including the system size,
	and the criteria for identifying surface states. Here we use parameters 
	$\lambda = 0.3, \mu=0.8 , t_1=1$ and ${\tilde t}_2=0.4$. We show the armchair edge spectrum in \subref{fig:subfig13},
	and the zigzag spectrum in \subref{fig:subfig12}. The surface states on the two sides of the system perfectly overlap in 
	\subref{fig:subfig13}.
	}
	\label{fig:clean_models_4}
\end{figure}

Here we explore an example of an alternative construction of a helical metal
model, without introducing new degrees of freedom. Both constructions were briefly outlined in 
Section~\ref{numerics}. We start with the Kane-Mele model\cite{Kane:2005A,Kane:2005B},
and add second neighbor hopping (which involves hopping only on the same sublattice).
This realizes the construction leading to a momentum-dependent chemical potential. 
We adjust the second neighbor hopping and overall chemical potential to be strong enough to close the bulk gap,
while having a near zero value near those momenta at which the surface states appear, on the zigzag edge surface of the honeycomb lattice.
We explore both this edge as well as the armchair edge. Our model is most simply and succinctly written as
\be\label{H_1}
{\mathcal H} = 
- t_1 \sum_{\langle i j \rangle \alpha} \left[a^{\dagger}_{i \alpha} b^{\phantom\dagger}_{j \alpha} + h.c. \right]
- \mu \sum_{j \alpha} a^{\dagger}_{j \alpha} a^{\phantom\dagger}_{j \alpha} + (a \rightarrow b)
+ \sum_{\langle \langle i j \rangle \rangle \alpha \beta} 
\left[
a^{\dagger}_{i \alpha} a^{\phantom\dagger}_{j \beta} \left( - {\tilde t}_2 \delta_{\alpha \beta} + i \lambda s^z_{\alpha \beta} \nu_{i j} \right) 
+ h.c. + \left( a \rightarrow b \right)  \right]
\; ,
\ee
where as in Eq.~\eqref{H_2}, $a,b$ denote the fermion operators on the two sublattices, 
$i,j$ denote the lattice sites, the spin indices are denoted by $\alpha,\beta$, 
and $\nu_{i j}$ is as defined in Refs.~\onlinecite{Kane:2005A,Kane:2005B} (and shown in Fig.~\ref{fig:model_conventions}).
Repeating the finite strip numerical diagonalization for both zigzag and 
armchair edges, we find the results of Fig.~\ref{fig:subfig12} and Fig.~\ref{fig:subfig13},
demonstrating yet again the presence of helical surface states coexisting with a bulk Fermi surface.
The surface states in the armchair edge geometry seem as if they have been pushed away from overlapping with the
bulk states, reminiscent of the exiling effects we discussed in the main text.

Note that Some examples of helical metals may also have surface states
on some faces but not on others, as would be the case in the above
model if the second-nearest neighbor hopping were sufficiently strong
such that no bulk gap would exist. Generically, such surface states will be
unstable to strong disorder, and will be subject to a finite life-time
as described in Sec. \ref{disorder} above. 

\subsection{Surface state multiplying}
\label{Three_states}

\begin{figure}
	\centering
		\subfigure[ Zigzag edge]{
			\includegraphics[width=3.0in]{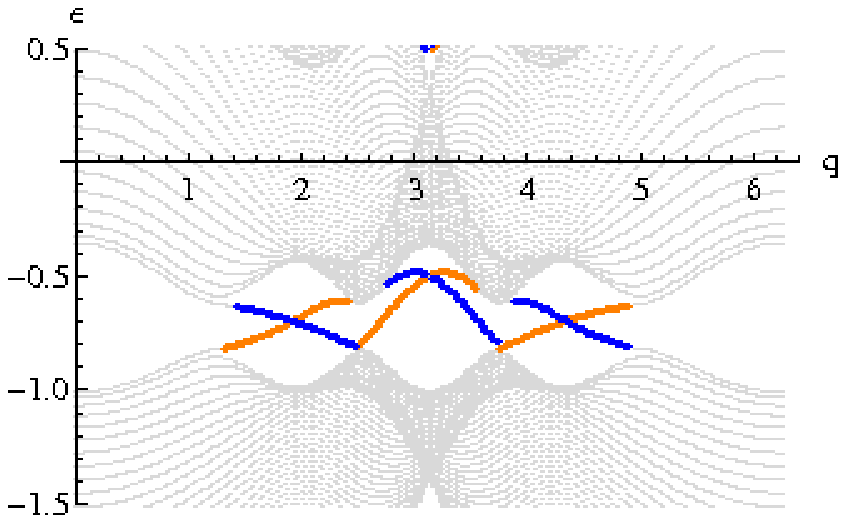}
			\label{fig:subfig10}
		}	
		\subfigure[ Armchair edge]{
			\includegraphics[width=3.0in]{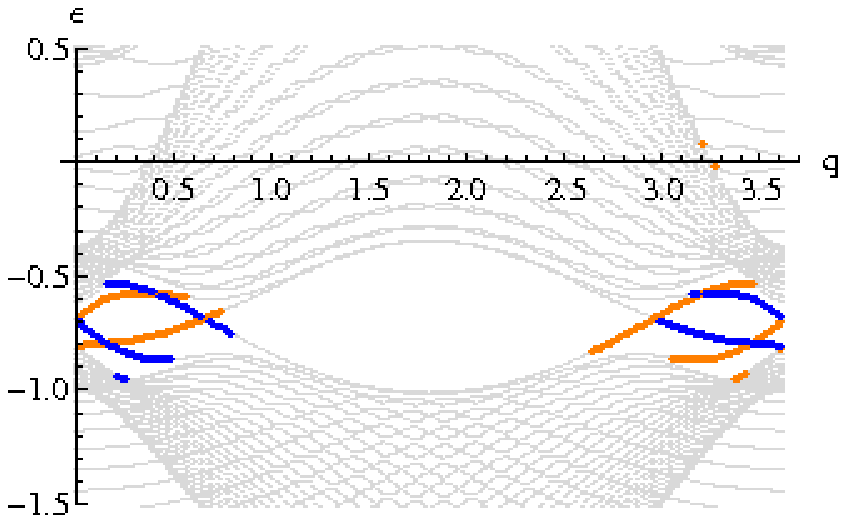}
			\label{fig:subfig11}
		}				
	\caption{
	All pictorial conventions are exactly the same as in Fig.~\ref{fig:3band_numerics}, including the system size,
	and the criteria for identifying surface states. Here we use parameters 
	$\lambda = 0.08,t_2=-1,t_1=1,h=2$ and $t_3=0.5$. We show the spectrum of both zigzag and armchair edge strip geometries.
	The same expulsion of the surface states occurs, as in Fig.~\ref{fig:3band_numerics}, and in addition we find 
	the number of surface branches has tripled to 3 Kramer's pairs on each wall of the strip.
	}
	\label{fig:clean_models_5}
	\end{figure}

Another revealing example of the model in Eq.~\eqref{H_2},
shows the helical nature of the surface states being preserved despite significant changes in the band structure.
The parameters used here in this example are $\lambda = 0.08,t_2=-1,t_1=1,h=2$ and $t_3=0.5$,
and we, once again, repeat the same numerical calculation on both the zigzag and armchair edged strip geometry.
The various spectra are displayed in Fig.~\ref{fig:clean_models_5}
and show that on both edges, the single surface state branch has multiplied into 3 co-propagating surface state branches,
albeit not crossing the Fermi energy (though that can be changed with the overall chemical potential)
so that the total number of helical surface states on one edge has changed from one Kramer's pair of bands to \emph{three}.
There are still an odd number of surface state branch pairs, which means the system has the same helical classification. 
This numerical result again confirms the robustness of the helical nature of the surface states in this model, 
despite strong mixing with an ordinary metallic band. 
We will leave the exploration of the edge-state
multiplication effect to future work.

\bibliographystyle{apsrev} 
\bibliography{Topo_metal_biblio}

\end{document}